\documentclass{JINST}
\usepackage{graphicx}
\usepackage{amsmath}   
\usepackage{subfigure}
\usepackage{wasysym}
\usepackage{lineno}
\usepackage{cite}

\title{Experimental study of electric breakdowns in liquid argon at centimeter scale}

\author{A.~Blatter, A.~Ereditato, C.-C.~Hsu, S.~Janos, I.~Kreslo\thanks{Corresponding author.}, M.~Luethi, C.~Rudolf~von~Rohr, M.~Schenk,  T.~Strauss,  M.~S.~Weber and M.~Zeller\\
\llap{}Albert Einstein Center for fundamental Physics, Laboratory for High Energy Physics, University of Bern,\\
Sidlerstrasse 5,3012 Bern, Switzerland\\
  E-mail: \email{igor.kreslo@lhep.unibe.ch}}

\abstract{In this paper we present results on measurements of the dielectric strength of liquid argon near its boiling point and cathode-anode distances in the range of 0.1~mm to 40~mm with spherical cathode and plane anode. We show that at such distances the applied electric field at which breakdowns occur is as low as 40~kV/cm. Flash-overs across the ribbed dielectric of the high voltage feed-through are observed for a length of 300~mm starting from a voltage of 55~kV. These results contribute to set reference for the breakdown-free design of ionization detectors, such as Liquid Argon Time Projection Chambers  (LAr TPC).}


\begin{document}

\section{Introduction}
Liquefied noble gases, such as argon or xenon, are becoming more and more popular media for ionization particle detectors to be employed in high energy  physics \cite{ATLAS}, neutrino physics \cite{rubbia77,ICARUS,wanf,GLACIER,Argoneut,uboone,LBNE}, dark matter search experiments \cite{ArDM, WARP, XENON,EXO} and other applications in particle and astroparticle physics. The required target (detector) mass gradually increases with the precision of the experiments. Large detectors, like Time Projection Chambers (TPC), in turn require high voltages in order to operate at reasonable drift field intensities. A significant amount of research activities has already been devoted for this subject (see for instance \cite{Badertscher:2012dq, Horikawa:2010bv, HVFT,ARGONTUBE0,ARGONTUBE1}).

Currently, the maximum stable breakdown-free operation voltage obtained in a liquid argon TPC (LAr TPC) was reported to be 100~kV in the 5~m long drift ARGONTUBE detector  that is operational in Bern \cite{ARGONTUBE2}. The oxygen-equivalent concentration of electronegative impurities achieved in ARGONTUBE is about 10$^{-10}$ (0.1~ppb). With ARGONTUBE we detected for the first time minimum ionizing particle (MIP) tracks across a drift length of about 5~m and with a signal-to-noise ratio of 16. However, one could not reach the ultimate design cathode-anode voltage of 500~kV because of high voltage breakdowns occurring in the medium outside of the drift volume at electric field intensities of about 30~kV/cm (evaluated by finite element numerical calculation). 
This value is substantially lower than the
commonly used reference value of 1.4~MV/cm for the dielectric strength of liquid argon, which was first measured and reported in \cite{SWAN60,SWAN61}. These authors showed a dependence of this value on the cathode-anode distance and elaborated a theoretical basis for its interpretation. However, the studies reported in \cite{SWAN60,SWAN61} were limited to cathode-anode distances of tens of microns and to small radius electrode geometry. Calculations also had such short distances as an assumption, allowing to simplify the equations and to yield to analytical solution. The purity of the argon used in those experiments was characterized as "spectroscopically pure", equivalent to an impurity of  $\approx$2$\times$10$^{-7}$ (0.2~ppm) oxygen equivalent.  

Some of the envisioned large scale neutrino detectors (e.g. \cite{GLACIER}) could require even longer drift distances than ARGONTUBE, up to tens of meters, hence implying voltages of the order of mega-volts. In order to yield a  sufficiently high collected charge for such long drift lengths, the argon must be purified to a concentration of oxygen-equivalent impurities lower than 0.1~ppb.
Therefore, a detailed knowledge of the behavior of the dielectric strength of ultra-pure liquefied noble gases (in particular argon) as a function of the cathode-anode distance and of the voltage is necessary for the design of the detectors.  

\section{Experimental setup}

\begin{figure}
\centering	
\includegraphics[width=0.265\linewidth]{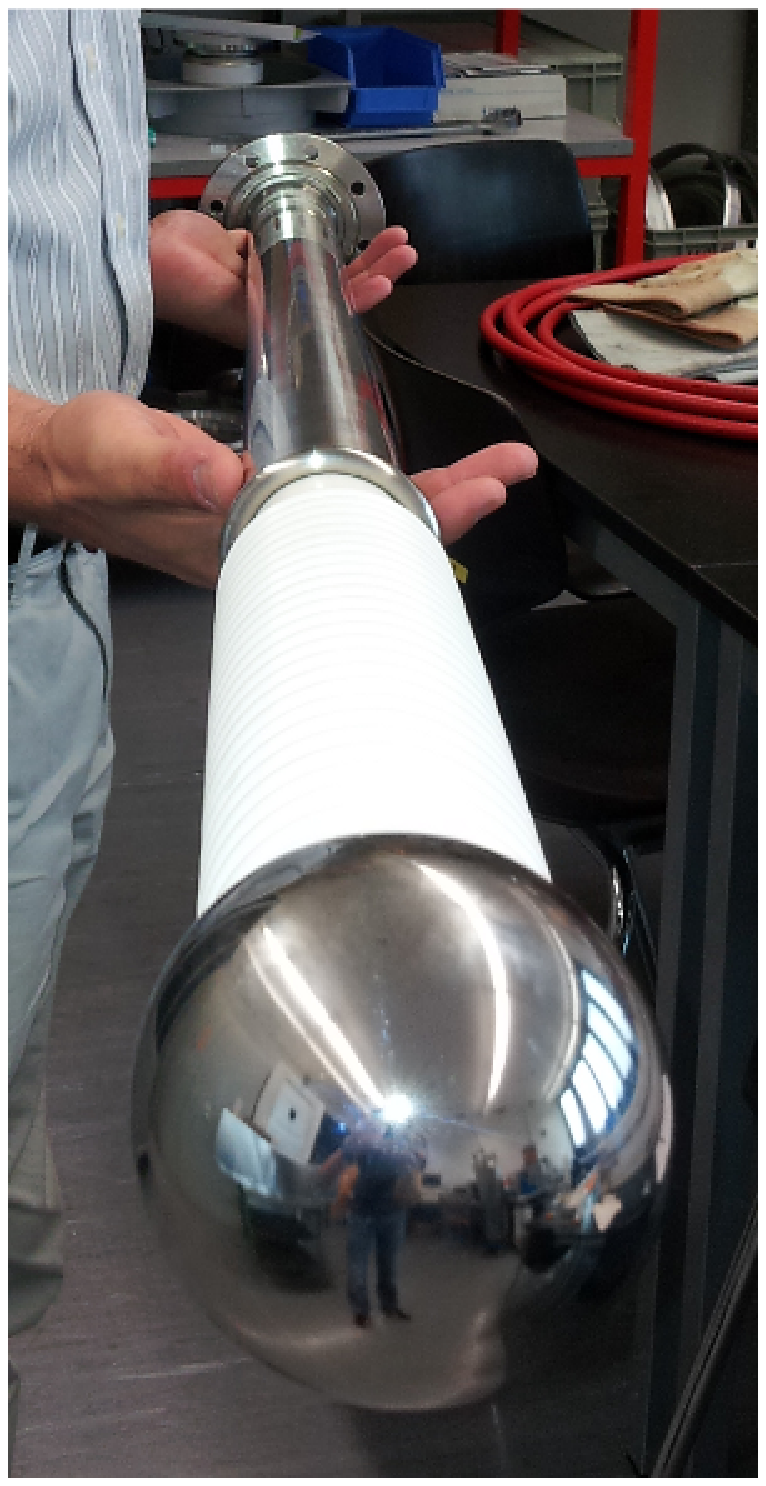}
\includegraphics[width=0.39\linewidth]{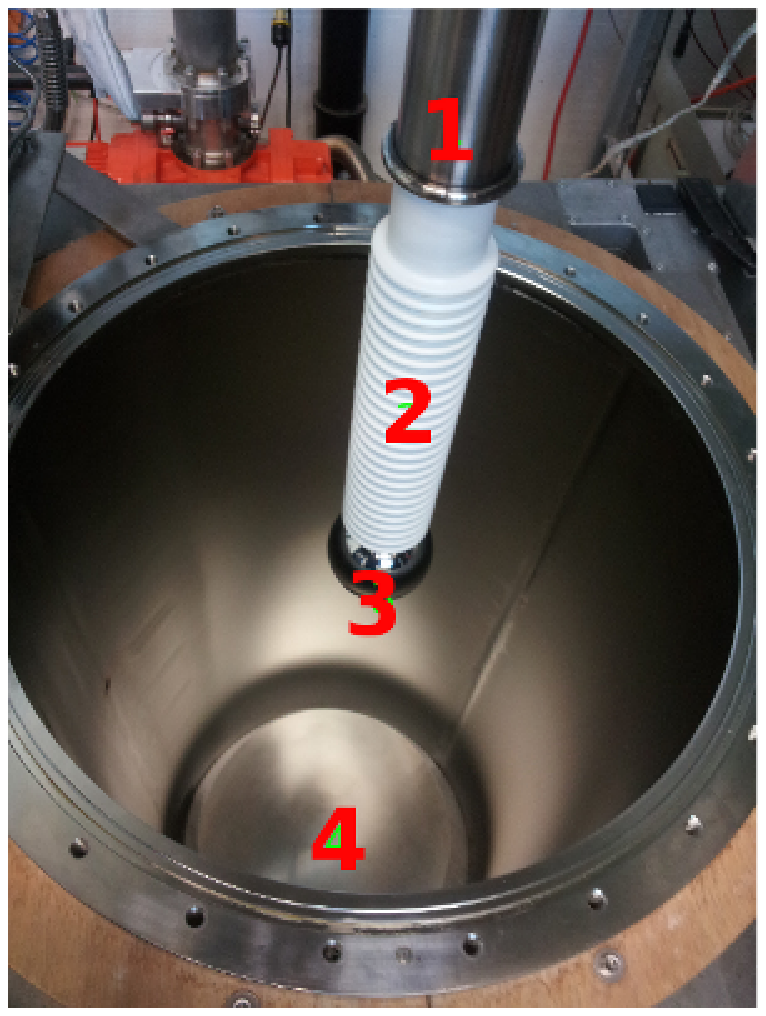}
\includegraphics[width=0.295\linewidth]{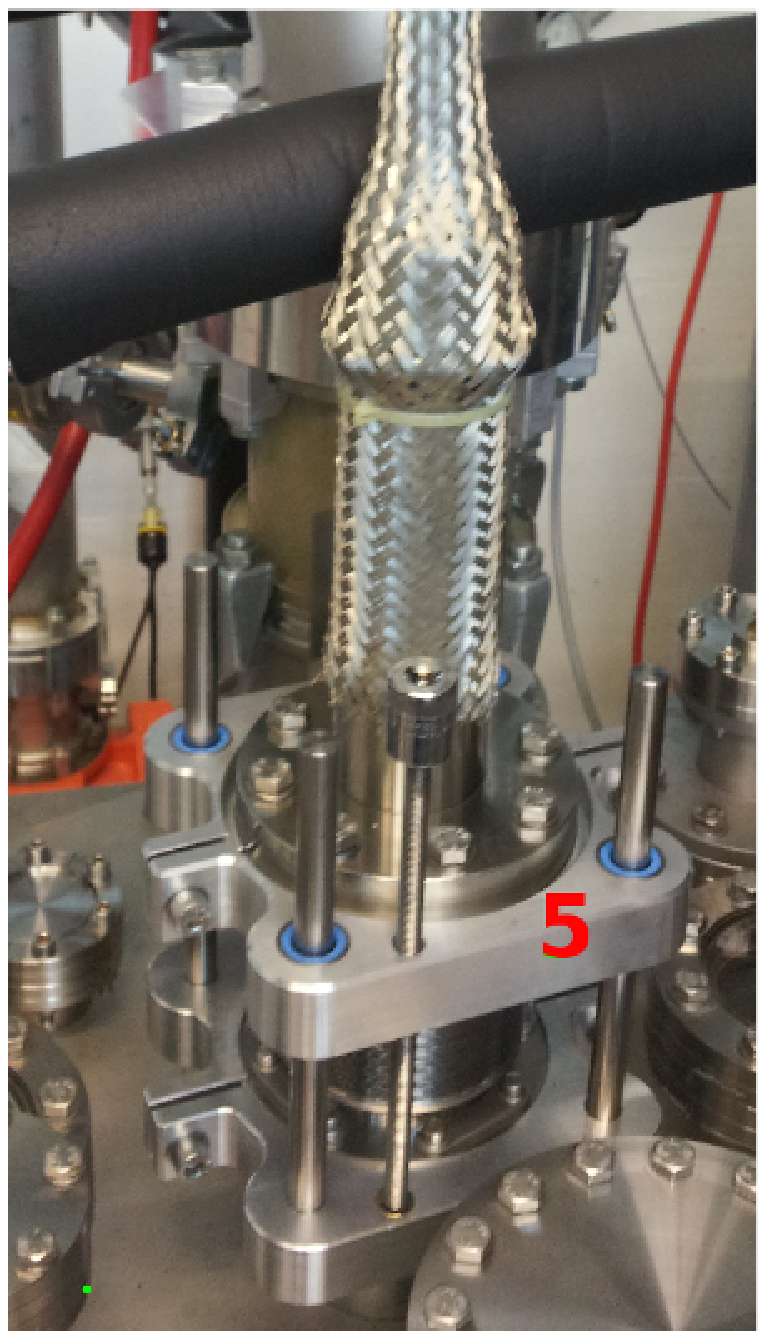}
\caption{Experimental setup. Left: high voltage feed-through with the spherical cathode. Middle: the feed-through before insertion into the cryostat. Right: linear translation unit used to set the cathode-anode gap width. 1- ground shield of the feed-through, 2- ribbed PET dielectric, 3-cathode sphere, 4- grounded bottom of the cryostat serving as anode.}
\label{fig:setup1}
\end{figure}

The setup we used to study the electric breakdown in liquid argon was hosted in a double-bath vacuum-insulated cryostat with the outer bath opened to atmosphere (Figure \ref{fig:setup1}). The inner bath was evacuated to $10^{-4}$ mbar and then filled with ultra-pure argon up to a level of about 60 cm. The level of the liquid argon in the outer bath was kept at about 40 cm above the inner one, to allow to close the inner volume after filling, in order to keep high liquid purity. The argon filled in the inner volume was purified by a pair of Oxysorb-Hydrosorb filters, which trap water and oxygen as well as other electronegative impurities.
The purity of argon after filling was estimated with a small time projection chamber according to the method described in \cite{Badhrees:2010zz} to be of the order of 1~ppb of oxygen-equivalent impurity concentration.

The pressure in the inner vessel was regulated by an overpressure valve in the range from 990~mbar to 1300~mbar. This allowed us to conduct the measurements
below the boiling point of liquid argon, excluding the formation of gas bubbles near the discharge gap.

\begin{figure}
\centering	
\includegraphics[width=0.8\linewidth]{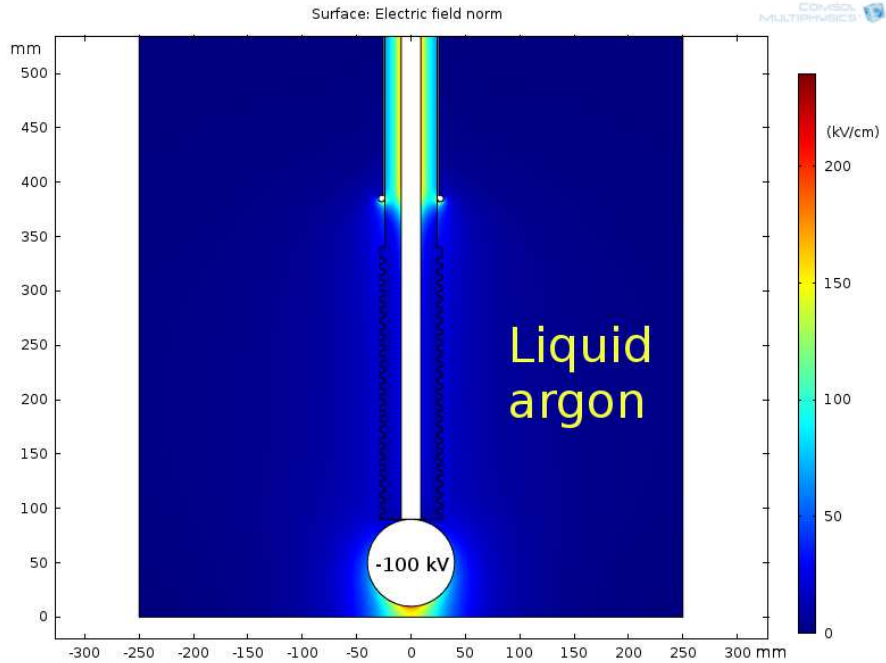}
\caption{Calculated electric field-amplitude map for the test setup with -100~kV at the cathode and cathode-anode distance of 1~cm.}
\label{fig:efield}
\end{figure}

The discharge gap was created by a cathode sphere of 80~mm diameter placed at a precisely measured distance from a flat anode plate at the bottom of the cryostat. Both electrodes were manufactured of stainless steel and mechanically polished to a clean mirror state. The cathode sphere was fixed at the 
high-voltage feed-through structure that provides mechanical support and feeds potentials up to -130~kV through the cryostat top flange into liquid argon to the cathode (see Figure \ref{fig:setup1}, left, middle). The feed-through was made using PET-C\footnote{Angst+Pfister AG CH-8052 Z\"urich.} polymer as a dielectric. The feed-through was mounted at the top flange with the aid of a translation unit (see Figure \ref{fig:setup1}, right) which provided vertical displacement in the range of 100~mm. The displacement was measured with an accuracy of
10~$\mu$m by a digital slide gauge. The electric field intensity in the discharge gap had its maximum at the cathode spherical surface at the point in front of the anode plate (Figure \ref{fig:efield}) in the full range of distances from 0 to 100~mm. Two quartz view-ports were installed on the top flange in order to visually observe breakdown sparks. 

A negative high-voltage potential was supplied to the feed-through central electrode from a Spellman SL130-150 power supply computer controlled via a dedicated control unit. This unit also provided instantaneous values of the output voltage and current back to the computer. A 20~MOhm current-limiting resistor was added to the circuit in order to limit the energy released in the discharge and
to minimize possible electrode surface damages. In addition, the output current of the power supply was limited to 50~$\mu$A.

\begin{figure}
\centering	
\includegraphics[width=0.99\linewidth]{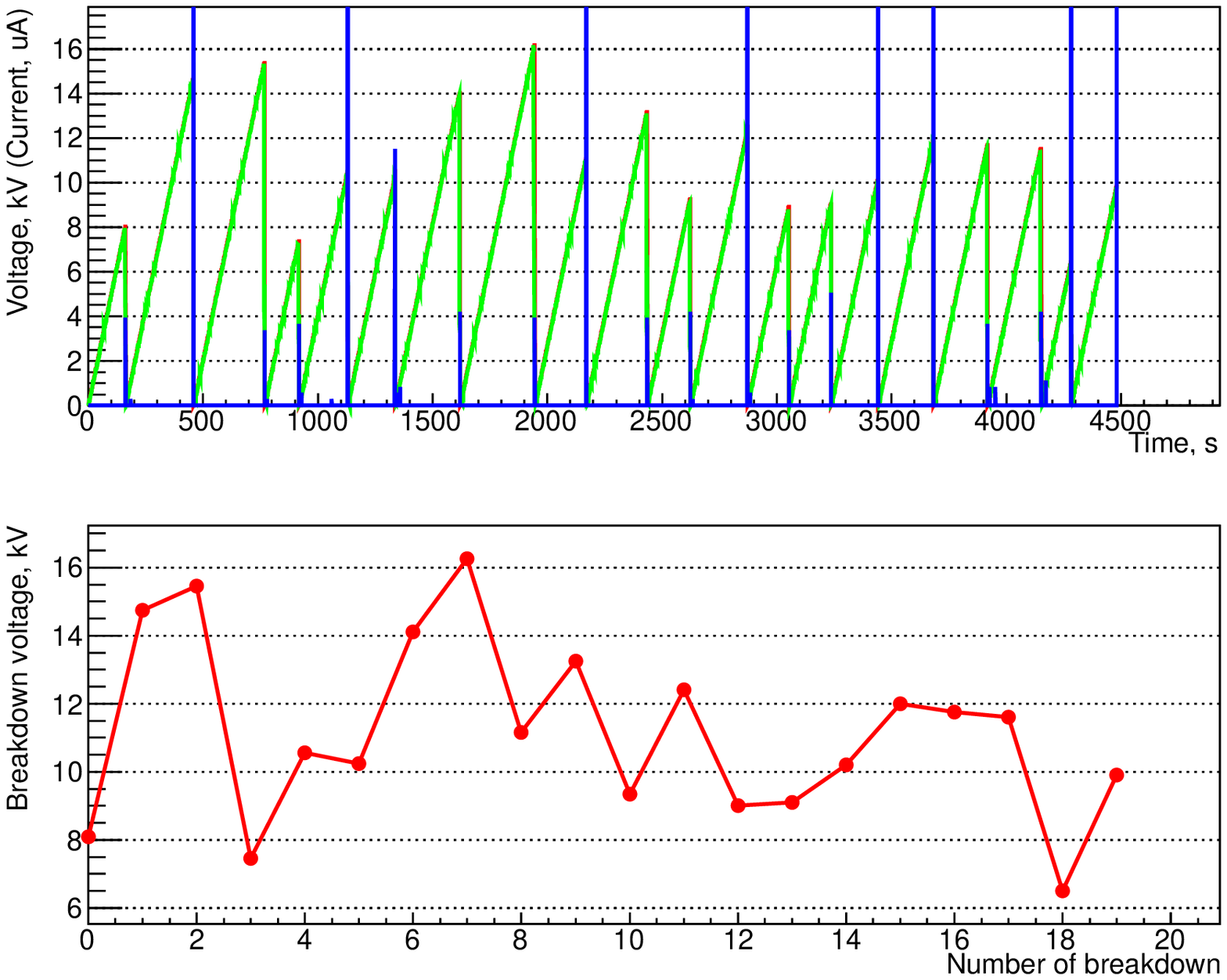}
\caption{Top plot: measured output voltage (green) and measured output current (blue). The 5~s pause at the end of each cycle is omitted from the graph. Bottom plot: registered breakdown voltage vs discharge sequential number.}
\label{fig:vset}
\end{figure}

The rate of ramping up the voltage was chosen to be low enough to decouple from possible dynamic charge-redistribution effects, while keeping the total measurement time within a reasonable range.
The voltage was raising from zero with a rate of 50 V/s up to the moment when cathode-anode spark discharge was registered by the power supply output current peak and voltage drop. At this moment, the highest reached voltage was stored as the breakdown voltage. The voltage was then lowered back to zero and kept so for 5~s in order to stabilize the setup and prepare for the next measurement cycle.
A typical graph illustrating such measurement sequence is shown in Figure \ref{fig:vset}.

\section{Measurement results}

The setup was put in operation at the end of 2013. Several thousand breakdown events in liquid argon were registered during a period of several days. To decouple from a possible deterioration of the argon purity, discharges for each cathode-anode distance were measured several times at the beginning, at the middle and at the end of the measurement period. No detectable variation was observed, allowing us to conclude that the argon purity did not reduce to a level that could affect the dielectric strength.

On the basis of a visual observation of the breakdown sparks, two types of breakdowns could be identified. The first (type 1) is a breakdown that occurs in the gap between the cathode sphere and the anode plate and is accompanied by light emitted directly from the gap below the sphere. At voltages higher than 55~kV, a second type (type 2) of breakdown starts to appear and even dominate for gap widths larger than 5-7~mm. These events are characterized by a flash-over at the surface of the feed-through ribbed dielectric. This second type of discharge was somehow unexpected. The distance between the cathode sphere and the ground electrode of the feed-through is 300~mm, much larger than the normal spark gap. The dielectric is manufactured with ribs in such a way that the total surface length is more than 600~mm. The configuration of the electric field at the surface of the dielectric is discouraging any kind of surface discharge, and the value of the electric field is much lower than that in the normal discharge gap.
 
\begin{figure}
\centering	
\includegraphics[width=0.8\linewidth]{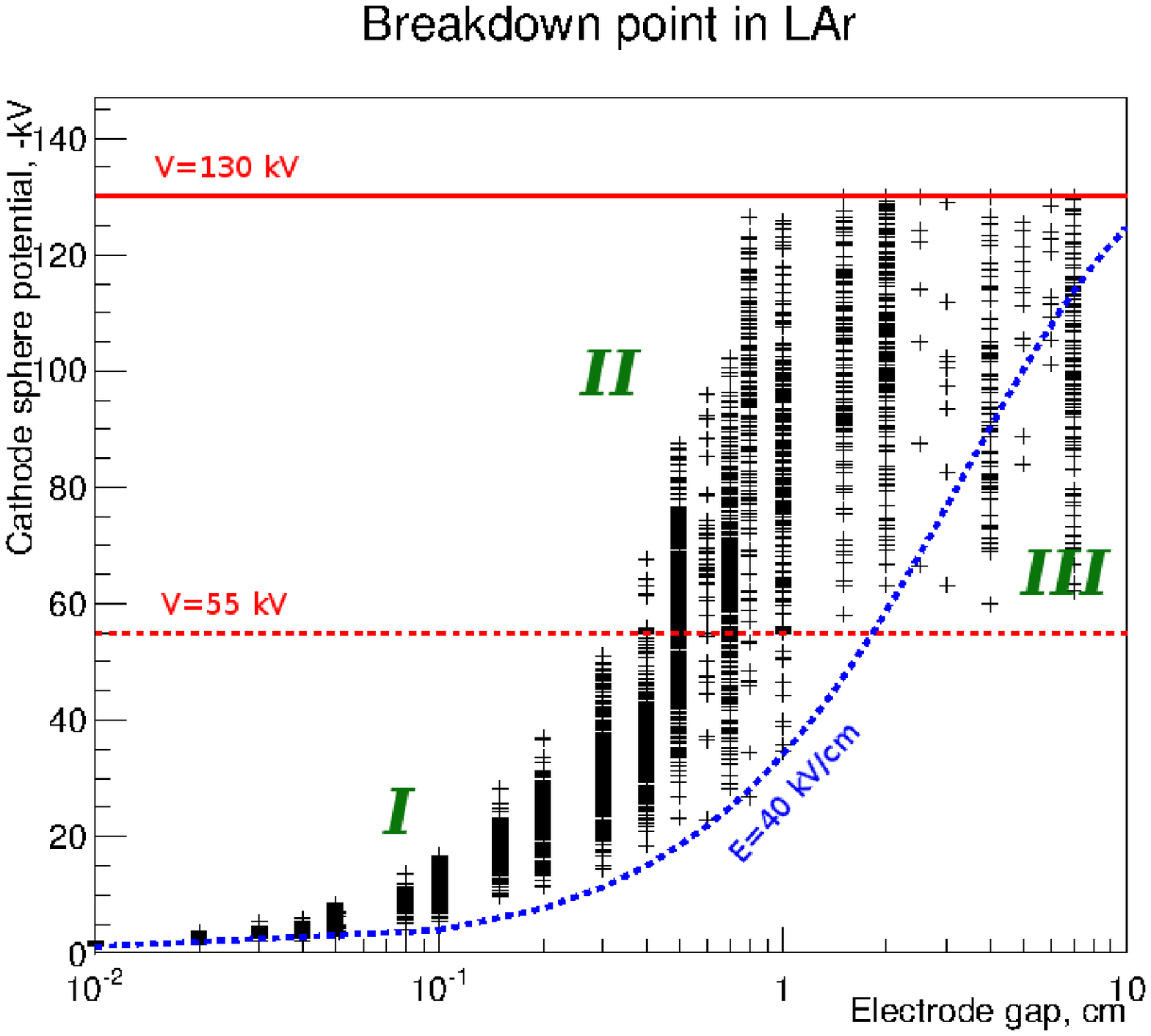}
\caption{Breakdown voltage vs cathode-anode distance. Solid red line: 130~kV - limitation of the power supply. The dashed red line marks the voltage of 55~kV. The blue line marks the field of 40~kV/cm on the sphere surface at the point of maximum. Zone~I - type~1 normal sphere-to-plane discharges are favorable; Zone~III - the voltage is high enough to trigger type~2 flash-overs, Zone~II - intermediate regime, where discharges of both types occur.}
\label{fig:summary}
\end{figure}

Each point of the plot of Figure \ref{fig:summary} corresponds to one discharge of type 1 or 2 in coordinates of breakdown voltage, as a function of the distance between the cathode sphere and the anode plate.  Zone I in the Figure corresponds to a relatively low voltage where type 1 discharges dominate. As the gap increases and the cathode potential exceeds 55 kV (dashed red line), flash-overs (type 2) start to appear (zone II) and finally take over (zone III). In the latter zone, the applied voltage is not high enough to break the gap, so the flash-over becomes more favorable and the dependence on the gap width vanishes. The solid red line denotes the maximum voltage of the power supply.

\begin{figure}
\centering	
\includegraphics[width=0.8\linewidth]{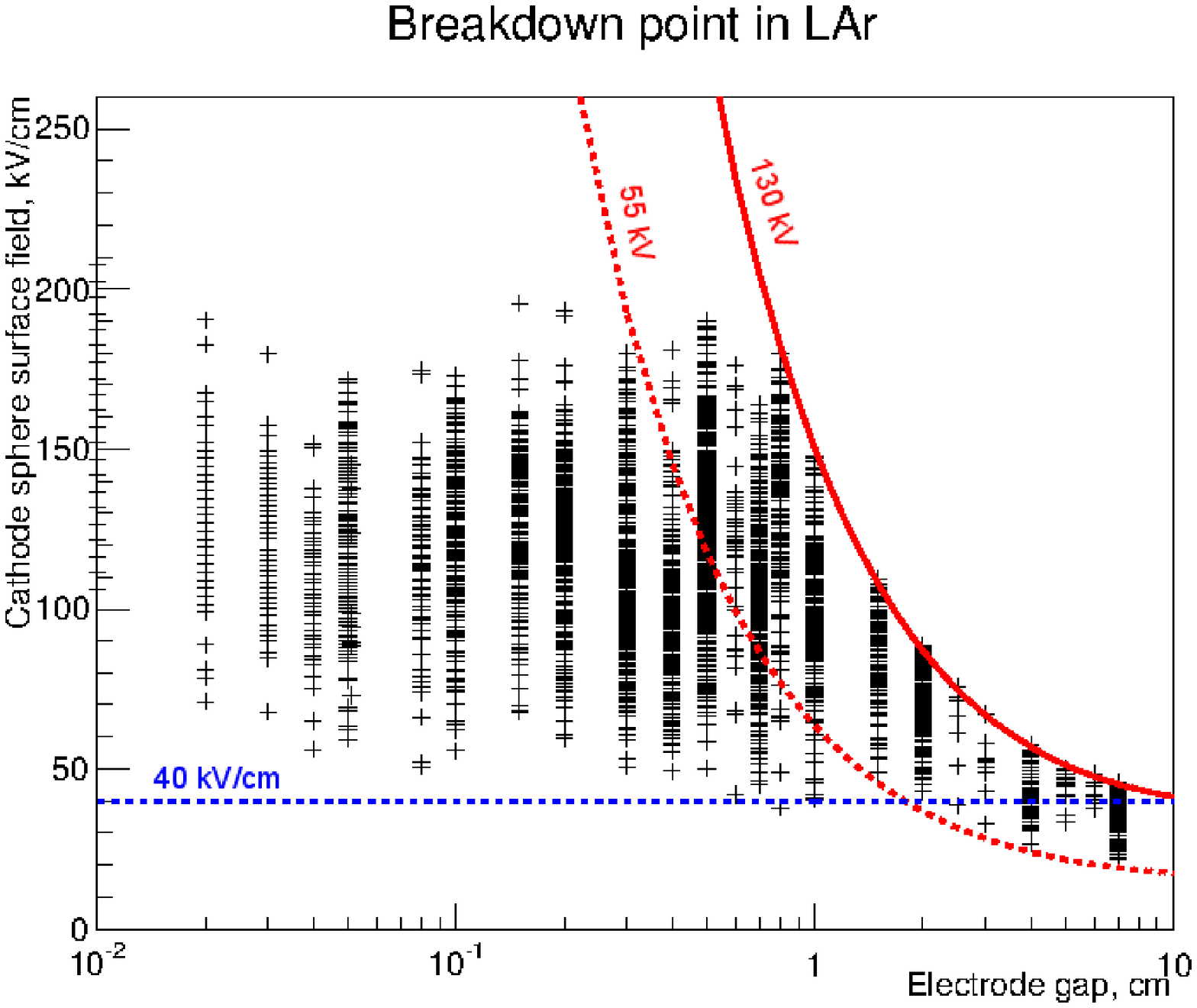}
\caption{Electric field at the cathode sphere surface at the breakdown vs cathode-anode distance. Solid red line: 130~kV - limitation of the power supply; the dashed red line marks the voltage of 55 kV; the blue line marks the field of 40~kV/cm.}
\label{fig:summary2}
\end{figure}

Figure \ref{fig:summary2} shows the same data plotted in coordinates of maximum electric field at the cathode sphere. This field was determined by finite element simulation with the Comsol Multiphysics package. The red lines mark the same values as in Figure \ref{fig:summary}; the blue line shows a field intensity of 40 kV/cm on the sphere surface at the point of its maximum. 

\section{Interpretation of the results}
The physics of electric breakdown in liquid dielectrics has been studied in some depth, mostly in relation to their use as insulators in electric power generation and transmission. The most studied dielectrics are therefore various types of hydrocarbons and oils. There are several reviews published on this topic (see e.g. \cite{review,russian}). Liquefied noble gases have been comparatively less studied. However, some models for initiators of the breakdown exist, based on space charge formation, gas bubbles and solid foreign particles \cite{polish}. Because of the ultra-high purity of the argon used in our experiments we can consider that only space charge and gas bubbles play an active role in the initiation of discharges in our conditions.

\begin{figure}
\centering	
\includegraphics[width=0.8\linewidth]{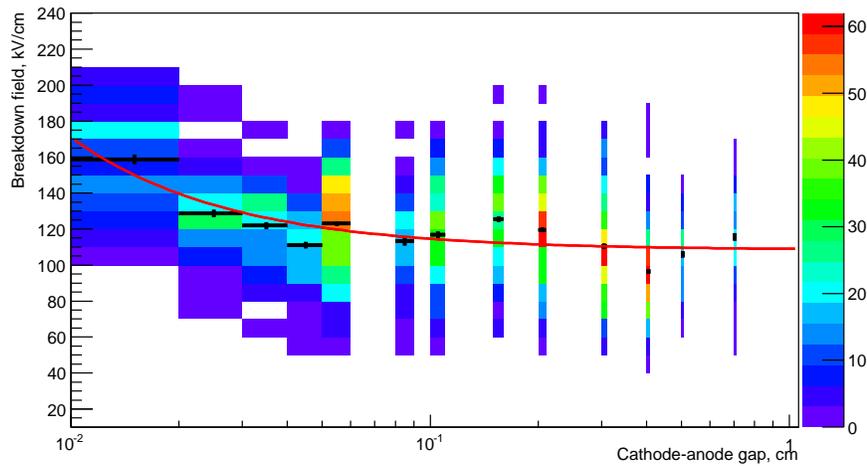}
\caption{Dielectric strength of liquid argon as a function of cathode-anode separation. The color scale shows the number of discharges per bin. The black crosses are placed at the average values of the breakdown field, showing the corresponding statistical errors. The red line shows fit with A+B/L where A=108.4~kV/cm and B=0.63~kV . }
\label{fig:type1}
\end{figure}

{\bf Type 1 discharge (sphere-plane gap).}
The region of the "normal" gap discharges (gap width $\le$7~mm) is shown in more detail in Figure \ref{fig:type1}.
In addition to the fact that at this scale the safe field intensity limit seems to be about 40 kV/cm, there is a 
noticeable drop of the average breakdown field with increasing gap width. This dependence was first observed by Swan in \cite{SWANDISTANCE}.
He described the mechanism of enhancement of the local cathode field due to accumulation of $Ar^+$ ions produced
in impact ionization by the drifting electrons. The main source of electrons in his model is given by field emission from the cathode surface. The first Townsend coefficient drives ion production in liquid argon. Breakdowns happen when the ion space charge that enhances field emission creates a positive feedback in the system. 
The analytical calculations by Swan were made assuming a relatively low field enhancement (<10\%) and short distance between the electrodes.
Qualitatively, increasing the gap width leads to higher ion production in the avalanche. Therefore, the value of the applied voltage required to reach instability and discharge decreases. These calculations reproduce the experimental measurements in the range of 10-100 $\mu$m in the sphere-sphere geometry with sphere radii of 2.5~mm.

For the measurements reported here the value of the applied electric field resulting in a breakdown is almost two orders of magnitude lower than that reported by Swan. At such low field intensity the first Townsend avalanche multiplication coefficient is very low: 
$\alpha \ll 1$~cm$^{-1}$ (see e.g. \cite{AtrazhevIakubov}). However, the low multiplication factor can be compensated by the long distance available for exponential avalanche development in ultra-pure argon with free electron life time of the order of milliseconds. This may create conditions for the production of high concentration of positive ion space charge in the volume and later at the cathode surface, eventually creating positive feedback.
In addition to that, a process similar to the Malter effect in gas ionization chambers \cite{Malter} was recently reported \cite{Zep,Brush,Forster,Huuckstadt}. These authors demonstrate the occurrence of spontaneous formation of (multi-)atomic layers of high order at the metal surfaces that behave like a solid crystalline material and 
therefore form an additional thin dielectric barrier to the drifting positive ions. Under these conditions, positive space charge
may create very high field values at the cathode surface, strongly enhancing electron field emission, which in turn can result into a discharge. 
Similar effects, in principle, take place in various hydrocarbons, for which an empiric dependence of the breakdown field on the electrode separation $L$ is suggested in \cite{russian}: $E_{br}=A+B/L$. Such interpolation is shown as a red line in Figure \ref{fig:type1}. The parameters of the fit: $A=$108.4~kV/cm and $B=$0.63~kV are in the same scale as reported in \cite{russian} (Table~13, p~61).

\begin{figure}
\centering	
\includegraphics[width=0.95\linewidth]{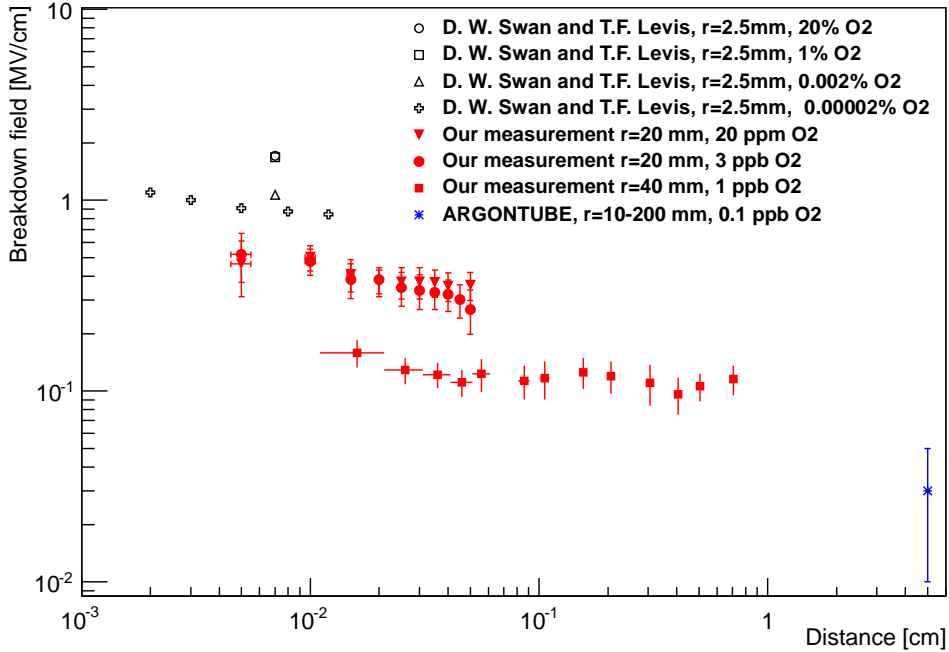}
\caption{Compilation of the experimental data on the electric strength of liquid argon including results from our measurements. 
}
\label{fig:all}
\end{figure}

\begin{figure}
\centering	
\includegraphics[width=0.8\linewidth]{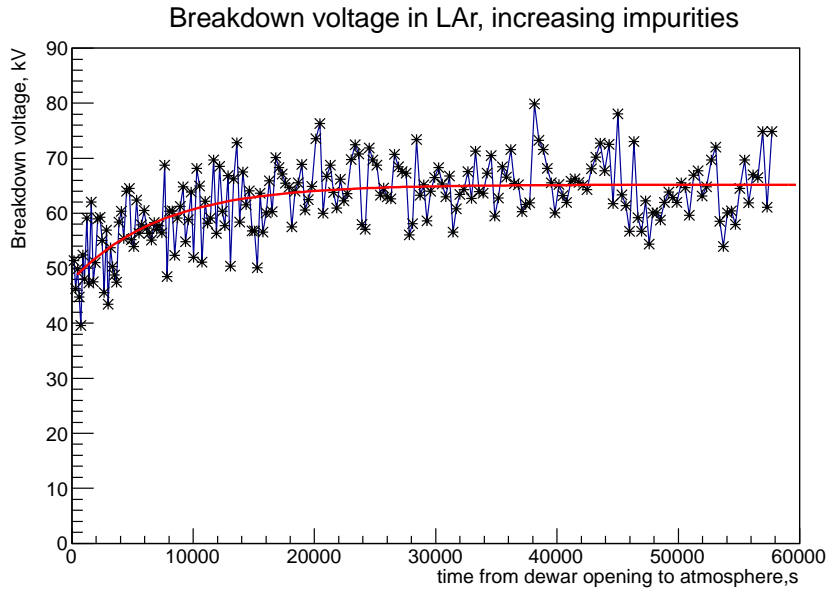}
\caption{Effect of increasing impurity concentration on the breakdown voltage for a gap width of 5~mm. The initial concentration is
1~ppb oxygen-equivalent impurities. At t=0 the dewar was opened to the atmosphere.}
\label{fig:pur}
\end{figure}

In the plot of Figure \ref{fig:all} we included the experimental data from \cite{SWAN60,SWAN61} and from our measurements with
two different geometries: sphere-sphere with R=20~mm and plane-sphere with R=40~mm. Data were taken with different values of argon purity and electrode radii (in \cite{SWAN60,SWAN61} R=2.5~mm). The influence of the
purity can be inferred from the data reported in Figure \ref{fig:pur}. The breakdown field increases by about 40\% from 1~ppb impurity concentration to equilibrium with atmosphere (> 1~ppm). The small effect of the purity is also confirmed by the fact that the two groups of points for 3~ppb and 20~ppm and R=20~mm in Figure \ref{fig:all} coincide within the error. This indicates that for the data in Figure \ref{fig:all} for R=20~mm and R=40~mm difference in purity (1~ppb to 3~ppb) can not explain alone the difference of a factor of three in the breakdown field. The cathode radius, on the other hand, may significantly affect the positive ion space charge density at the cathode surface and, in turn, the breakdown voltage. 

The hypothesis of a positive role of the ion space charge partly explains the large spread of the breakdown voltages for each gap width.
The flux of cosmic ray particles brings random contributions to the total charge produced in the high field intensity region.
Possible charge deposition induced by high-energy electromagnetic showers may cause breakdowns even at low initial fields. The probability 
to have such events in the critical volume rises with the scale of the setup and may contribute to the explanation of the large difference in the measured minimal breakdown field for micron-wide and centimeter-wide gaps. Precise analytical treatment of the underlaying physics at the scale of centimeters is by far not trivial and will be the subject of further studies. 

\begin{figure}[ht]
\centering	
\includegraphics[width=0.8\linewidth]{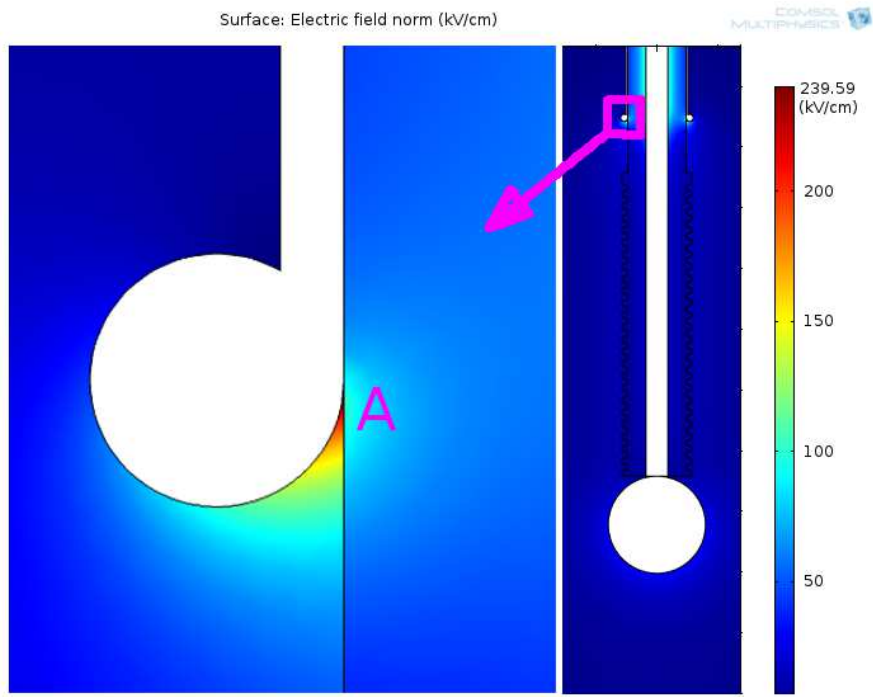}
\caption{Simulated electric field intensity near the ground electrode of the feed-through. At the point A the electric field value reaches a maximum of about 240 kV/cm at -100 kV cathode voltage.}
\label{fig:emax}
\end{figure}

The possibility of the initiation of type 1 discharges by formation of gas bubbles is ruled out by keeping the pressure in the inner vessel 
at 100~mbar above atmospheric pressure. The outer bath is opened to atmosphere, thus keeping the inner vessel temperature constant well below the boiling point. The liquid was still during the test and showed no boiling anywhere near the discharge gap region. The only region where seldom (less than one per ten seconds) vapor bubbles formation was observed is the lower edge of the feed-through ground shield (near point A in Figure \ref{fig:emax}), 35 cm of liquid above the discharge gap.

\begin{figure}[ht]
\centering	
\includegraphics[width=0.35\linewidth]{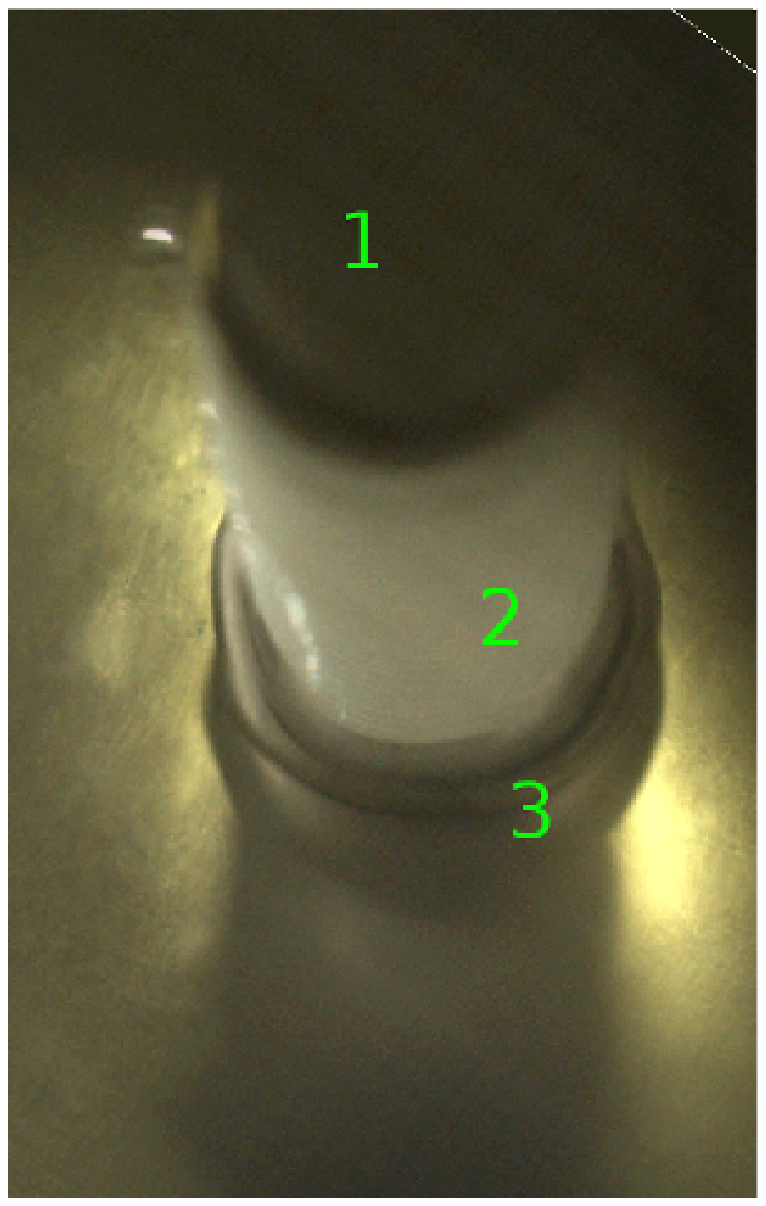}
\includegraphics[width=0.35\linewidth]{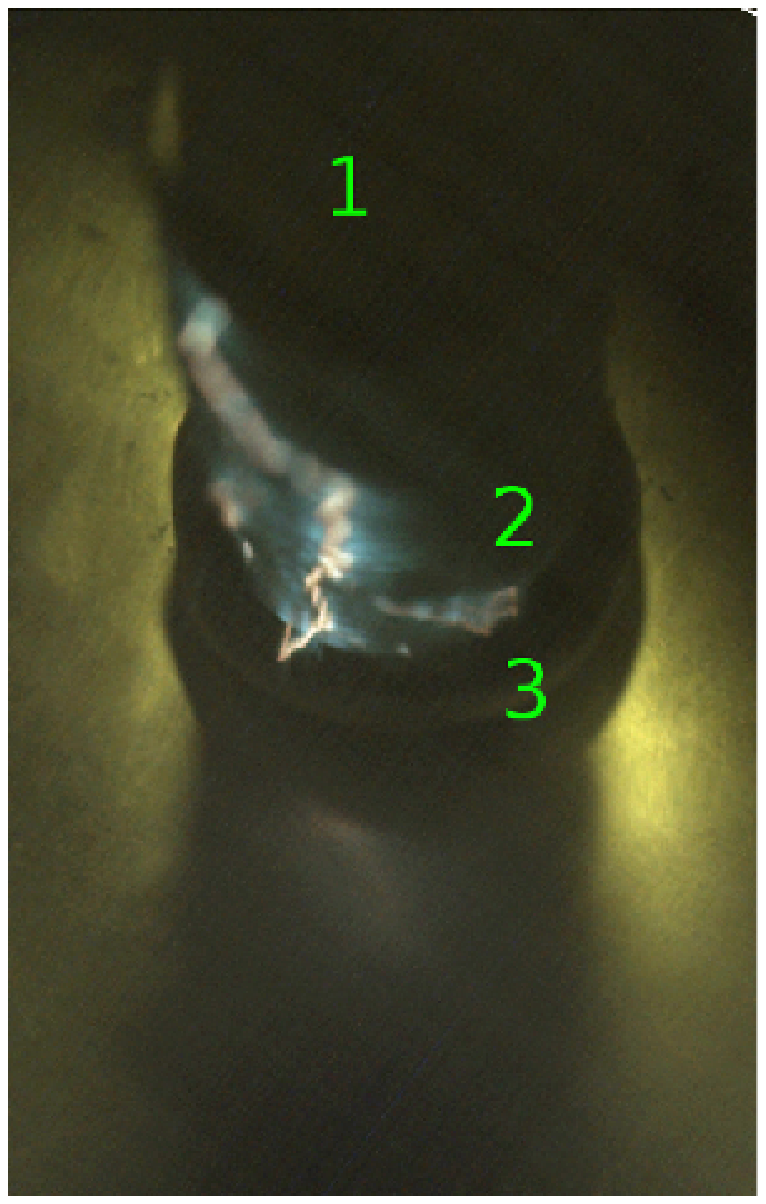}
\caption{Type 2 discharge. Left: Initiation of discharge in a bubble (left top corner); Right: flash-over filaments 4~ms later. 1- ground shield of the feed-through, 2- ribbed PET dielectric, 3-cathode sphere.}
\label{fig:glow}
\end{figure}

{\bf Type 2 discharge (flash-over).}
In spite of the experimental configuration being very unfavorable for the flash-over and low value of the electric field at the surface of the PET feed-through dielectric, such flash-over may be provoked by an initial discharge at the feed-through ground shield (point A in Figure \ref{fig:emax}).
Although very local, this field is higher than the one on the sphere-plane gap. Moreover, due to the heat flux from the top of the
feed-through, this point is most favorable for the appearance of the gas phase (bubbles). In fact, we have observed suppression of this type of discharges when we raised the level of the liquid argon to improve cooling of the feed-through and thus reduce bubble production. This allowed to extend type 1 discharge zone up to 7~mm gap width. It seems that once initiated in this region, a flash-over develops towards the cathode across large distances, even in conditions of very low (less than 20~kV/cm) electric field (see Figure \ref{fig:glow}, left).
The discharge filament pattern differs from spark to spark. Therefore, the breakdown does not result in formation of a burned-in preferential surface path.
Most likely, breakdown filaments jump from rib to rib, supported by an accumulated volume charge between ribs rather than following the surface of the dielectric (Figure \ref{fig:glow}, right). In favor of this hypothesis is the fact that in the vast majority of the observed discharges there are multiple filaments. To clarify this issue a very fast high-resolution image sequence recording will be required. This will be the subject for our further study. 

\section{Conclusions}

Early results from measurements of liquid argon dielectric strength conducted in the 1960's indicated values as large as 1.4~MV/cm. However, already at that time, mechanisms that may lead to a significant dependence of this value on the cathode-anode distance were considered. The lack of experimental measurements at a scale larger than a few hundred microns has significantly retarded the further development of the studies. The measurements presented in this paper extend the scale of cathode-anode gap widths up to 10~mm with spherical cathode and plane anode. It is shown that for such scale electric breakdowns in highly purified liquid argon (1~ppb oxygen equivalent) occur at field intensities significantly lower then expected, namely as low as 40~kV/cm. The observation of a dependence of the breakdown field on the cathode-anode distance and on the cathode shape supports the hypothesis that the breakdown is governed by space-charge effects in the volume and at the cathode surface. Further extension of the explored scale up to 10~cm, as well as the theoretical interpretation of the data will be required to set reference for the design of future large ionization detectors based on the liquid argon technology.

\section{Acknowledgement}
The authors would like to thank S.~Delaquis for his important contribution to the design of the feed-through used for the experimental setup. This work was funded by the grant 200020-149246 of the Swiss National Science Foundation and by a grant of the Canton of Bern.


\end{document}